\documentstyle[prl,aps,twocolumn]{revtex}
\include{epsf}
\epsfverbosetrue
\begin{document}
\twocolumn[\hsize\textwidth\columnwidth\hsize\csname @twocolumnfalse\endcsname

\draft

\title{Surface shape resonances in lamellar metallic gratings}

\author{T. L\'opez-Rios$^1$, D. Mendoza$^1$, F.J. Garc\'{\i}a-Vidal$^2$,
J. S\'anchez-Dehesa$^2$, and B. Pannetier$^3$} 

\address{$^1$Laboratoire d'\'Etudes des propiet\'es Electroniques des Solides-CNRS, BP166\\
 38042 Grenoble Cedex 9. France.} 
\address{
$^2$ Departamento de F\'{\i}sica Te\'orica de la Materia Condensada.
 Facultad de Ciencias(C-V),\\ Universidad Aut\'onoma de Madrid. E-28049 Madrid. Spain.} 

\address{$^3$Centre de Recherches sur les Tres Basses Temperatures-CNRS, BP166\\ 
38042 Grenoble Cedex 9. France.}

\date{\today} 
\maketitle

\begin{abstract}

The specular reflectivity of lamellar gratings of gold with grooves 0.5 microns wide
separated by a distance of 3.5 microns was measured on the 2000 cm$^{-1}$ - 7000 cm$^{-1}$
spectral range for p-polarized light. For the first time, experimental evidence of the 
excitation of electromagnetic surface shape resonances for optical frequencies is given. 
In these resonances the electric field is highly localized inside 
the grooves and is almost zero in all other regions. For grooves 
of depth equal to 0.6 microns, we have analyzed one of these modes  
whose wavelength (3.3 microns) is much greater 
than the lateral dimension of the grooves.

\end{abstract}
 
]

\narrowtext

Metallic gratings can exhibit absorption anomalies \cite{Wood}.
 One of these anomalies which is particularly remarkable is observed 
for p-polarized light only, and is due to surface plasmon polariton (SPP) excitations.
 SPP excitation induces a minimum on the specular reflectance 
 spectra which is indicative of the amount of energy flowing parallel
 to the surface.
 In a first order approximation, the spectral position of the minimum does
 not depend on the shape or the amplitude of the grooves but 
 depends only on the dielectric constant and period of the grating.
 One interesting problem which has been 
raised long time ago and which is still of interest today is the near field 
dependence of these modes on the grating shape\cite{weber}.
 Linked to this problem is the possible existence
 of modes localized in grooves of prominent shape and their relation with 
non-linear optical effects observed in certain rough metal surfaces [3-8].

At the beginning of the century Rayleigh pointed out that flat rigid surfaces 
with cylindrical holes can present acoustic resonances for well defined depths 
of the holes\cite{ray}.
 Rayleigh showed that under these resonant conditions the 
acoustic energy is concentrated in the holes and he suggested that similar 
effects could occur with light.
More recently Rendell and Scalapino\cite{ren}
suggested the possible existence of localized plasmons 
 in order to
explain light emission in metal-oxide-metal structures.
 These plasmon 
modes are qualitatively different from propagative SPPs on a 
flat surface excited by attenuated total reflection or by a gentle 
 surface corrugation.
 Despite the conceptual and practical interest of these surface
 shape resonances and 
well documented theoretical predictions, till now no experimental evidence 
of these electromagnetic resonances has been reported for optical frequencies.

 In this letter we show that 
for lamellar gratings with deep rectangular cross sections, localized waveguide 
resonances which are equivalent to the acoustic resonances described by 
Rayleigh, can be excited in the channels when the impinging light has an 
electric field component perpendicular to the grooves direction.
The experiments here presented also illustrate the existence of hybrid modes,  
combination of standing waves localized in the grooves with
propagating SPPs. 

Measured samples consist in periodic arrays of metallic grooves of
nominal width 0.5 microns and separated 3.5 microns. The pattern
extends over an area 1$\times$1 cm$^2$. The samples were prepared on the surface of a
silicon wafer by standard photolithography. After development of a  
negative resist, the sample was etched in SF6 plasma down to the desired depth.
The resist mask was subsequently
removed by reactive ion etching in an oxygen plasma. Finally the structured
silicon surface was metallized by thermal evaporation of a gold layer. The
substrate was rotated during the evaporation in order to coat both the
bottom and the walls of the grooves and the average gold layer is about 100 nm thick.
Fig.1 shows a scanning electron
micrograph of a sample with grooves of depth 0.2 microns. 
Also in Fig. 1 we show a schematic picture of our samples with a 
definition of the different parameters: width of the 
grooves, $a=0.5$ $\mu m$, period of the grating, $d=3.5$ $\mu m$ and angle 
of incidence of p-polarized light, $\theta=21^0$.  
A Fourier transform spectrometer was used to measure the reflectance of 
these samples using a flat surface of gold as a reference. 

Firstly computational studies  
of scattering of plane waves by these lamellar metallic gratings were carried out.
Our aim was to analyze the appropriate values 
for the depths of the grooves that can support 
surface shape electromagnetic resonances.     
For that purpose we have used a {\it Transfer Matrix} formalism \cite{Pendry} which  
is ideally suited to work with materials like gold for which the dielectric
response disperses with frequency. 
For these calculations 
we have used the dielectric functions of gold 
as tabulated in Ref.\onlinecite{Palik}. 
Knowledge 
of the EM transfer matrix of the system allows us to calculate 
transmission and reflection coefficients for an incoming plane wave. 
Once these matrices are obtained the reflectance of the grating 
and a real-space picture of the resulting {\bf E}-field can be easily calculated.  

Fig. 2 displays computed values of the specular reflectance for p-polarized 
light and ange of incidence $\theta=21^0$ for grooves of 
different depths as a function of the wavenumber ($k$) of 
the incoming plane wave. 
We then show how for small values of the depths ($h=0.1-0.2$ $\mu m$), the reflectivity  
minima correspond to SPP excitations as they are
observed with an almost flat surface. The energetic positions of these modes are 
indicated by arrows in Fig. 2.
With increasing depth, the resonances are broadened due to radiation damping
and all reflectance minima become inequivalent; 
the spectral position of some of them ($k=2100$ cm$^{-1}$ and $k=4450$ cm$^{-1}$) changes only 
slightly with $h$ whereas the position of others (in particular the minimum at 
$k=4200$ cm$^{-1}$ for $h=0.1$ $\mu m$) depends strongly on the depth of the grooves.

In order to gain physical insight into this 
problem and explain the different features observed in Fig. 2, we have also 
carried out an approximated analysis of this scattering process. 
We take gold as a perfect metal and we assume that,  
as the wavelength of light is much greater than the width
of the grooves,
incident light only excites the fundamental eigenmode of the grooves \cite{wirgin1}.
The amplitude of this excited mode is proportional to $1/D$,
the denominator $D$ being:

\begin{eqnarray}
D & = & \cot(k h) - 
i \frac{a}{d} \sum_{n=-\infty}^{\infty} \frac{[{\rm sinc}(k\gamma_n a/2)]^2}{(1-\gamma_n^2)^{\frac{1}{2}}}
\end{eqnarray}
\noindent with ${\rm sinc}(\xi) \equiv \sin\xi/\xi$.
$\gamma_n=\sin\theta + \frac{ 2 \pi n}{k d}$
is associated with the $n$th reflected diffraction order.

The specular reflection coefficient $r_0$ can be written: 

\begin{eqnarray}
r_0 & = & 1 + 
2i \frac{a}{d} \frac{[{\rm sinc}(k\gamma_0 a/2)]^2}{(1-\gamma_0^2)^{\frac{1}{2}}} \frac{1}{D}
\end{eqnarray}
\noindent so the specular reflectance $R_0=|r_0|^2$ also depends on $D$.

The zeros in the real part of $D$ are linked to electromagnetic surface resonances; 
for these particular wavenumbers, the amplitude of the fundamental eigenmode 
of the grooves is maximum provoking    
a reduction in the specular reflectance (see Eq.(1)).
In Fig. 3 we show the energetic position of the zeros of Re($D$) for 
$a=0.5$ $\mu m$, $d=3.5$ $\mu m$ and $\theta=21^0$ as a function of 
the depth of the grooves ($h$). 
For small depths, the electromagnetic resonances are SPPs 
whose spectral positions can be calculated using 
simple diffraction theory. The wavenumber of these 
SPPs are given by the diffraction condition 
$\gamma_n=\pm 1$ (see Eq.(1)) and are located 
at $k=2100$ ($n=-1$) , $k=4200$ ($n=-2$), 
$k=4450$ ($n=2$) and $k=6300$ ($n=-3$) cm$^{-1}$ 
 in the frequency range used in our experiments. 
For larger depths, however, 
the excited modes have a hybrid character, combination of 
SPPs with waveguide modes localized at the grooves. The energetic positions 
of these waveguide modes for an {\it isolated} groove of depth 
$h$ are given by the simple relation $\cos(kh)=0$. When forming a periodic 
array, as it is in our case, the grooves interact electromagnetically and dispersion 
relations of these waveguide modes change slightly \cite{lopez} and are displayed by circles in 
Fig. 3. 
This figure clearly shows 
how the spectral locations of excited electromagnetic modes in deep metallic gratings (dots) 
are the result of hybridization of flat lines associated with SPPs (no 
dependence with $h$) 
and dispersion curves of waveguide modes (circles) that vary with $h$ as $1/h$. 
For example, the reflectance dip located at $2100$ cm$^{-1}$ 
for $h \rightarrow 0$ in Fig. 2 shows no dependence with $h$ for small depths because the 
hybridization of the SPP mode at this wavenumber with the corresponding waveguide mode  
occurs for depths larger than $0.6$-$0.7$ $\mu m$. On the contrary, SPP located at $4200$ cm$^{-1}$  
evolves rapidly with $h$ and for $h=0.4$ $\mu m$ its energetic 
position almost coincides with the position of the corresponding 
waveguide mode localized in the grooves.
This implies that this mode might have a predominant waveguide character.
In view of these results, lamellar metallic gratings of period $d=3.5$ $\mu m$ with 
rectangular cross section of width $a=0.5$ $\mu m$ and depth larger than 
$0.4$ $\mu m$ will have at least one surface shape electromagnetic resonance 
in the optical regime.
 
In order to observe this surface shape electromagnetic resonance we construct a 
metallic grating of depth equal to $0.6$ microns using the technique 
described above. For this particular value of $h$, 
the surface shape mode is located at around $3000$ cm$^{-1}$ and another 
electromagnetic mode located at $2100$ cm$^{-1}$ that in principle 
should have a predominant SPP character 
could be observed (see Fig. 3). 
In Fig. 4 we show the experimental specular 
reflectance of this sample as a function of the wavenumber of 
incoming p-polarized light. Also in this figure  
we present the corresponding theoretical calculation for this 
value of $h$ using 
our {\it Transfer Matrix} formalism.
Fig. 4 shows how 
incident light is exciting both electromagnetic modes in this 
metallic grating. The reflectance curve also presents a dip 
at around $4700$ cm$^{-1}$ that, as a difference to the dips 
located at lower energies, is not associated to a zero of Re($D$) but 
to a minimum of $|D|^2$.
The shape and spectral positions of the different  
dips are well described by our theoretical model. 
The experimental features are rounding off due to the 
angle dispersion of incident beam. 
The narrow resonance located at $2100$ cm$^{-1}$ presents a 
very high intensity of the {\bf E}-field at the upper corners 
of the grooves (around 300 times larger than intensity of incoming light).
Although its energetic location almost coincides with the 
spectral position of the SPP mode, this resonance 
already presents a hybrid character  
and is quite similar to sharp surface plasmon resonances recently 
observed in 
deep sinusoidal gratings \cite{wat}. 

Focussing our attention in the surface shape resonance located at around 
$3000$ cm$^{-1}$, it is interesting to analyze its evolution  
as a function of the depth of the grooves, $h$. In other words, 
how a delocalized SPP at $h \rightarrow 0$ evolves to form    
a localized waveguide mode for larger $h$. In Fig. 5 we show a detailed 
picture of the resulting electric fields for different values of $h$:  
(a) $h=0.2$ $\mu m$, (b) $h=0.4$ $\mu m$ 
and (c) $h=0.6$ $\mu m$ and wavenumbers, $k$, that correspond to the spectral 
locations of this surface shape resonance for the values of $h$ analyzed. 
For $h=0.2$ $\mu m$, the 
fields are very weak in the grooves and intense at the external surface as 
it is well known to occur for SPP's modes. 
With increasing $h$ ( $h=0.4$ $\mu m$), 
the electric field is entering in the grooves and the mode 
has a clear hybrid character.   
For $h=0.6$ $\mu m$ (the depth of the grooves in our sample), 
the intensity of the {\bf E}-field is mainly concentrated in the grooves 
and is practically zero 
in all other regions (vacuum or metal). The maximum intensity in the grooves is around 
100 times larger than the intensity of the incoming {\bf E}-field and hence 
at this frequency an enhanced infrared absorption selective to molecules chemisorbed in the grooves 
is expected. 
It is interesting to point out that the strength 
of the {\bf E}-field associated to these surface shape resonances is inversely proportional 
to the ratio between the width of the grooves ($a$) and period of the grating ($d$)  
(see Eq. (2)). Hence for channels of   
nanometric dimensions  (that could be made today with edge technology) 
extremely high fields can be achieved if the depth of the grooves is properly chosen. 
However, narrow channels naturally exit at the 
grain boundaries of poorly crystallized metallic films.
 Almost one century 
ago Wood\cite{wood2} observed optical absorption anomalies of coldly deposited 
alkali metals which was then attributed to light being trapped in the 
cavities existing in the metals. 
Our results suggest that cold deposited films could be 
particular systems where metallic cavities can support these localized 
surface shape resonances. 

In conclusion we have observed surface shape resonances for 
optical frequencies in lamellar metallic 
gratings with deep rectangular grooves of micron dimensions.  
For grooves of depth 0.6 microns, we have analyzed  
one of these resonances located at around $3000$ cm$^{-1}$ in 
which the intensity of the {\bf E}-field is highly localized 
inside the channels.
These modes are to some extent similar to the surface plasmons which
induces some kind of transparency of the metallic plates with holes of lateral
dimension much smaller than the wavelength \cite{ebb}.
The effects here described could also be at the origin of unusual enhancements
of infrared absorption of molecules on transparent metallic films \cite{yos}.
The experiments presented above suggest that 
composite media including metallic regions in close vicinity could give rise 
to extremely strong electromagnetic resonances. 

The authors would like to thank J. L. Tholence for his help to bring up the
collaboration leading to this work.
and  to Th. Crozes and Th. Fournier for help and advises in
sample preparation. We also wish to thank D. Lafont and D. Mariolle
(CEA-LETI Grenoble) for the SEM micrograph of the samples.
D. Mendoza acknowledges a sabbatical leave from UNAM , Mexico and financial support
 from this institution.
J. S\'anchez-Dehesa acknowledges financial support
from the Comisi\'on Interministerial de Ciencia y Tecnolog\'{\i}a, contract
MAT97-0698-C04.

{\Large {\bf Figure Captions}} \newline

{\bf Figure 1}. A scanning electron micrograph of a sample that 
consist in a metallic grating with grooves of depth $h=0.2$ $\mu m$. 
In the upper part of this figure we show a schematic view of our samples with 
grooves of width $a=0.5$ $\mu m$ and separated by $d=3.5$ $\mu m$. 
The incident light is p-polarized and the angle of incidence $\theta=21^0$. 

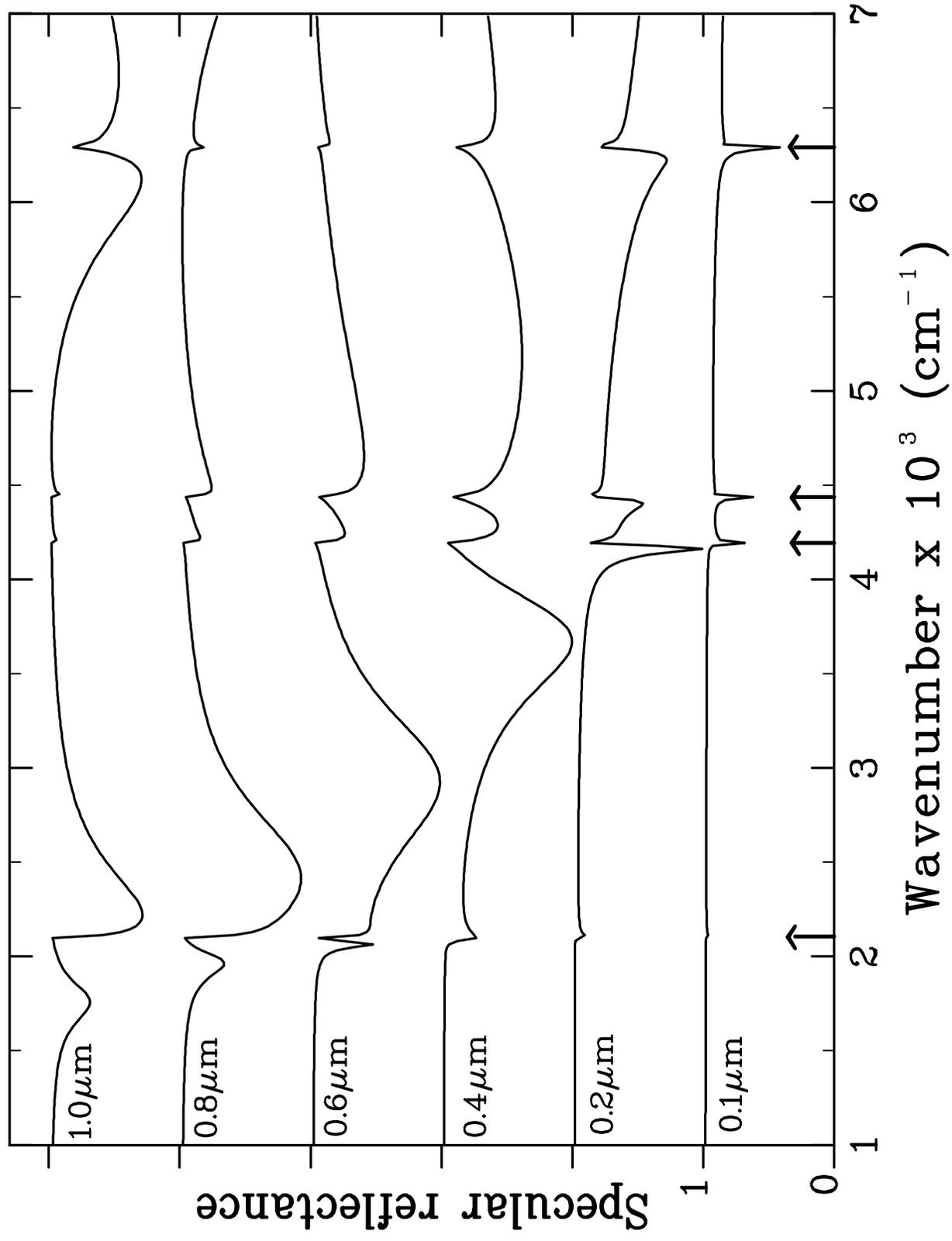
{\bf Figure 2}. Calculated specular reflectance of a gold grating 
with the parameters as defined in Fig.1. This magnitude is analyzed as 
a function of the wavenumber of the incident light  
for increasing values of the 
depth of the grooves going from $h=0.1-1.0$ $\mu m$. 
For the sake of clarity each curve is shifted up by $+1$ with 
respect to the previous one.

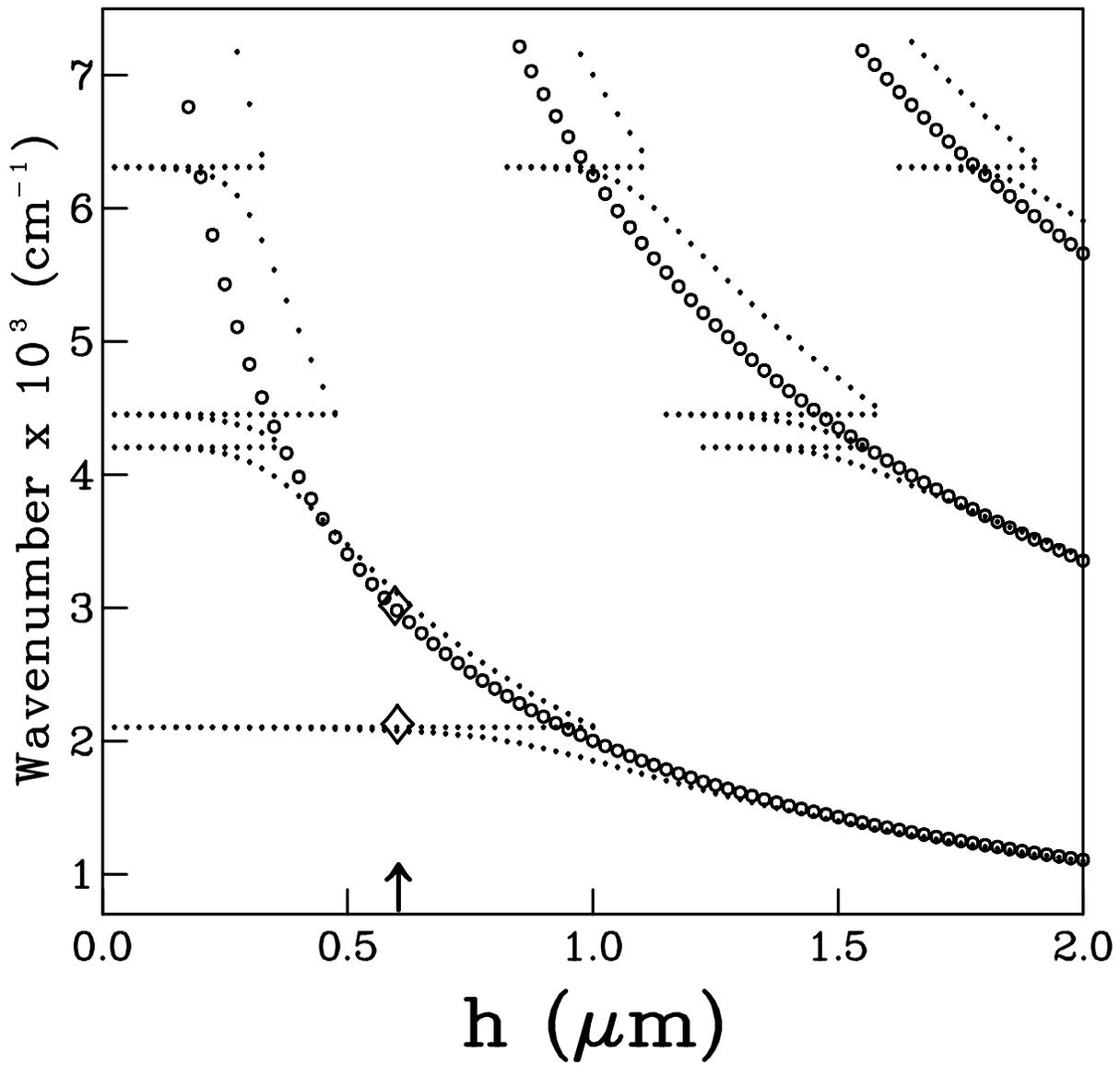
{\bf Figure 3}. Energetic positions of zeros of Re(D) (dots) and 
waveguide modes localized in the grooves (circles) for different 
values of the depth of the grooves, $h$. The locations 
of the excited modes for $h=0.6$ $\mu m$ are marked by diamonds 
in the figure. 

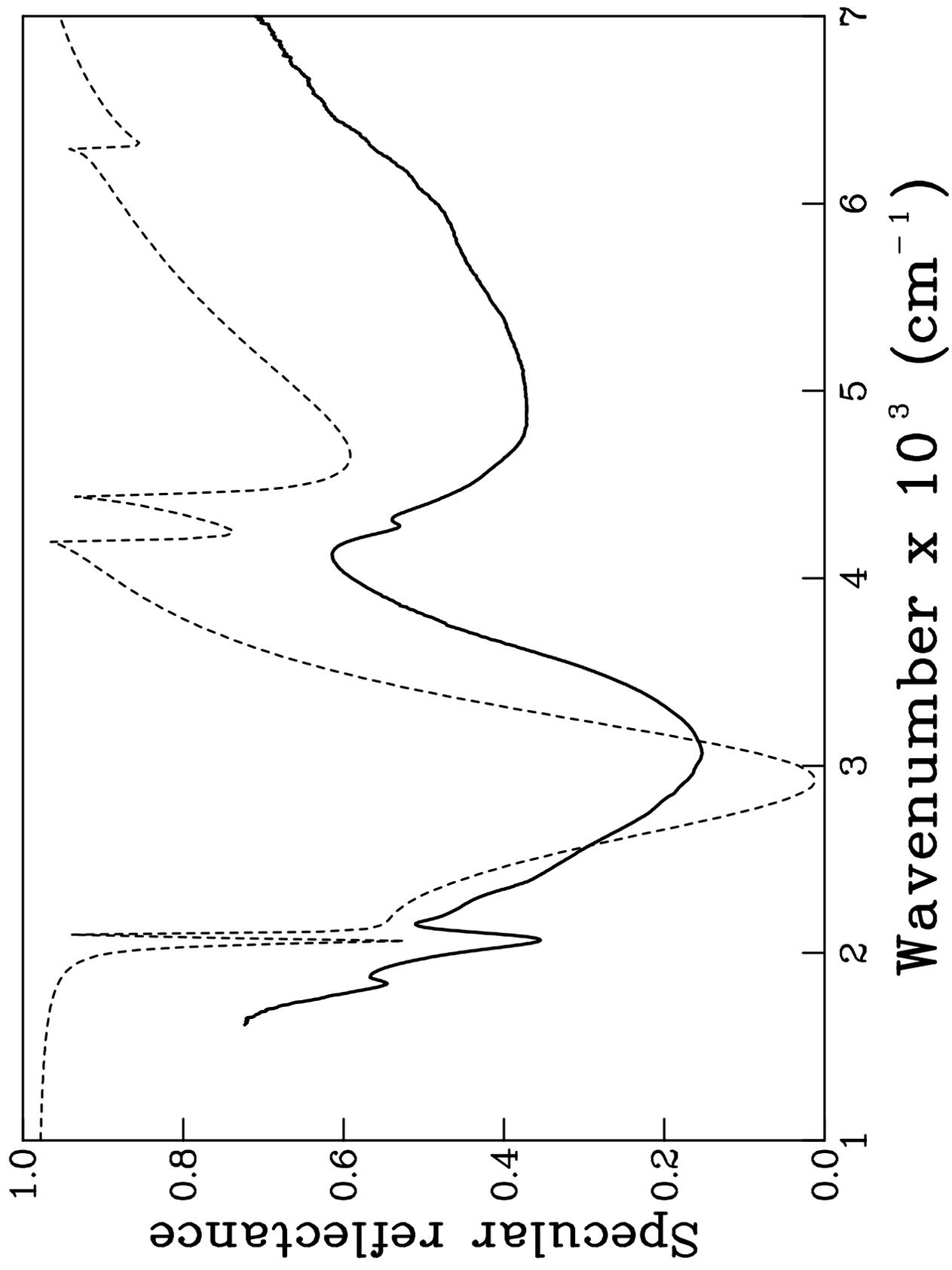
{\bf Figure 4}. Experimental (full line) and computed values (dashed line) of the 
specular reflectance of a gold grating with grooves of 
depth $h=0.6$ $\mu m$, period $a=3.5$ $\mu m$ and width $d=0.5$ $\mu m$.
The angle of incidence is $21^0$.  

{\bf Figure 5}. Detailed pictures of the intensity of the 
{\bf E}-fields in two unit cells 
of gold gratings with grooves 0.5 microns wide and 
3.5 microns separated. The shape of the grating is 
also drawn in the figure.  
The intensities are shown in a grey scale (white: minimum  
intensity, black: maximum intensity) for different values of 
$h$ and wavenumber of incident light:   
(a) $h=0.2$ $\mu m$ ($k=4160$ $cm^{-1}$), (b) $h=0.4$ $\mu m$ 
($k=3680$ $cm^{-1}$) and (c) $h=0.6$ $\mu m$ ($k=2920$ $cm^{-1}$). 
This last case corresponds to the excitation 
of a surface shape resonance. 

\begin{references}

\vspace*{-1.5cm}

\bibitem{Wood}
R. W. Wood, Proc. R. Soc. London A {\bf 18}, 269 (1902).

\bibitem{weber}
M. Weber and D.L. Mills, Phys. Rev. B {\bf 27}, 2698 (1983).

\bibitem{wirgin1}
A. Wirgin and T. L\'opez-R\'{\i}os,  Opt. Comm. {\bf 48}, 416 (1984);
{\it Erratum}, Opt. Comm. {\bf 49}, 455 (1984).

\bibitem{lopez}
T. L\'opez-R\'{\i}os and A. Wirgin , Solid State Comm. {\bf 52}, 197 (1984).

\bibitem{wirgin2}
A. Wirgin and A.A Maradudin, Phys. Rev. B {\bf 31}, 5573 (1985).

\bibitem{wirgin3}
A. Wirgin and A.A. Maradudin, Prog. Surf. Sci. {\bf 22}, 1 (1986).

\bibitem{garcia}
F.J. Garc\'{\i}a-Vidal and J.B. Pendry, Phys. Rev. Lett. {\bf 77}, 1163 (1996).

\bibitem{maradudin}
A.A. Maradudin, A.V. Shchegrov and T.A. Leskova, Opt. Comm. {\bf 135}, 352 (1997).

\bibitem{ray}
Lord Rayleigh, Phil. Mag. {\bf 39}, 225 (1920).

\bibitem{ren}
R. W. Rendell and D.J. Scalapino, Phys. Rev. B {\bf 24}, 3276 (1981).

\bibitem{Pendry}
J.B. Pendry, J. Mod. Opt. {\bf 41}, 209 (1994); P.M. Bell {\it et al.}, 
Comput. Phys. Comm. {\bf 85}, 306 (1995).

\bibitem{Palik}  {\it Handbook of Optical Constants of Solids }, edited by
E.D. Palik (Academic, Orlando) (1985).

\bibitem{wat}
R. A. Watts, T.W. Preist and J. R. Sambles, Phys. Rev. Lett. {\bf 79}, 3978 (1997).


\bibitem{wood2}
R. W. Wood, Phil.Mag. {\bf 38}, 98 ( 1919).

\bibitem{ebb}
T. W. Ebbesen, H. J. Lezec, H. F. Ghaemi, T. Thio and P.A. Wolff,  Nature {\bf 391}, 667 (1998).

\bibitem{yos}
T. Yoshidome, T. Inoue and S. Kamata, Chem. Lett.{\bf 1}, 533 (1997).

\end{references}
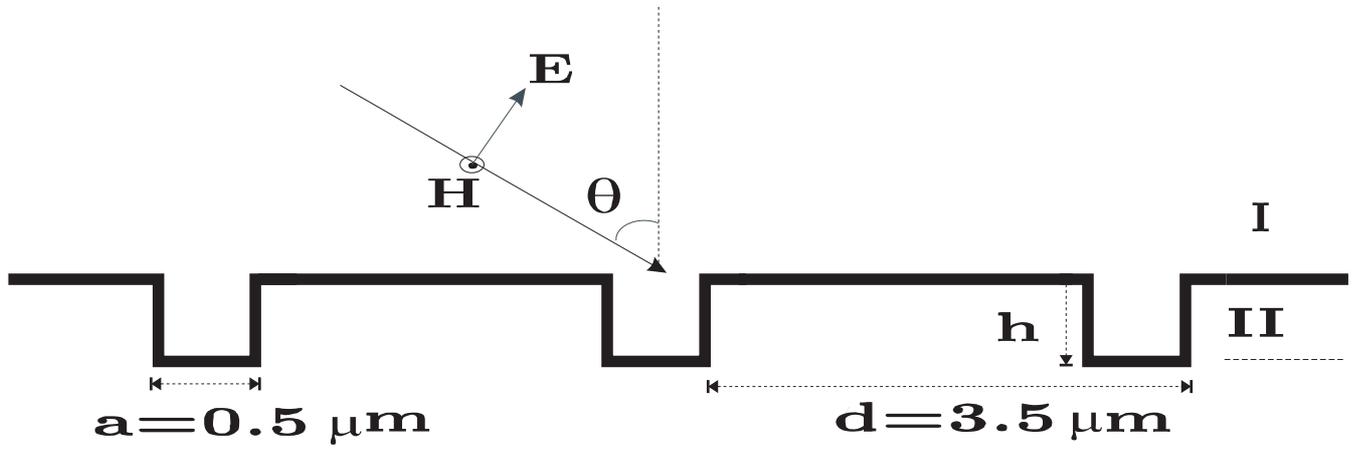
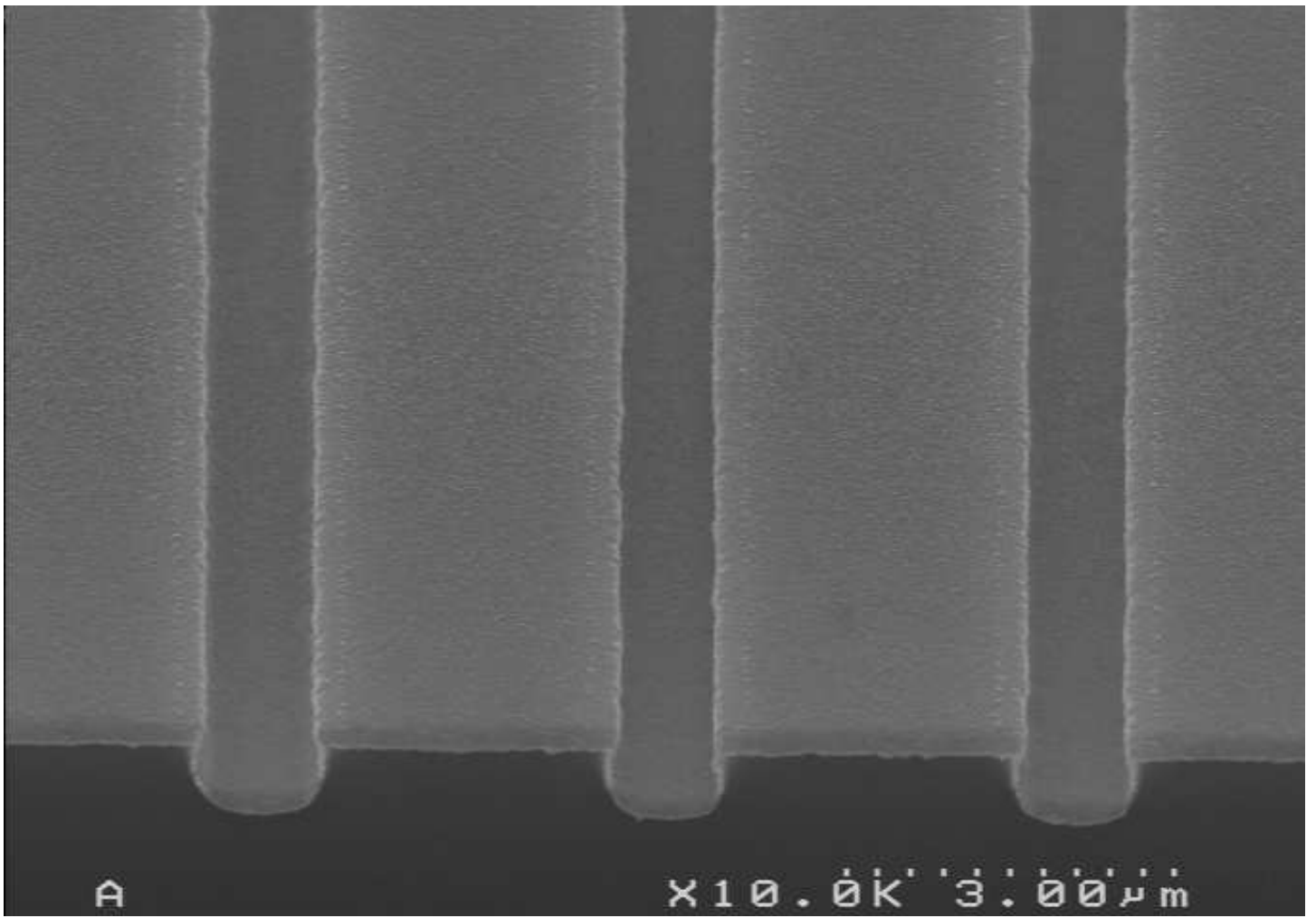
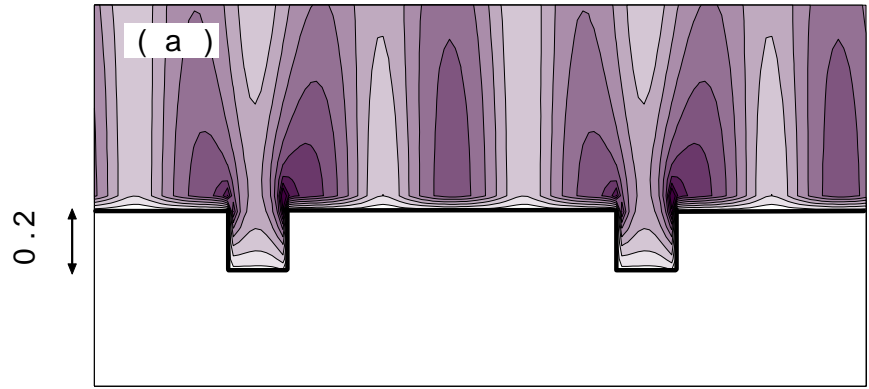
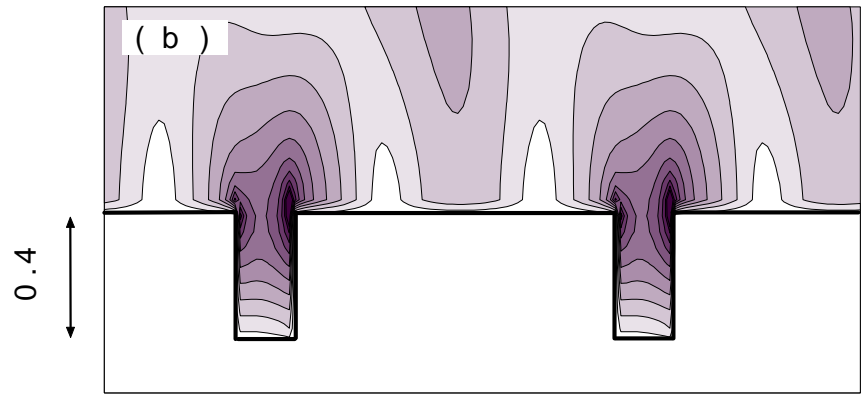
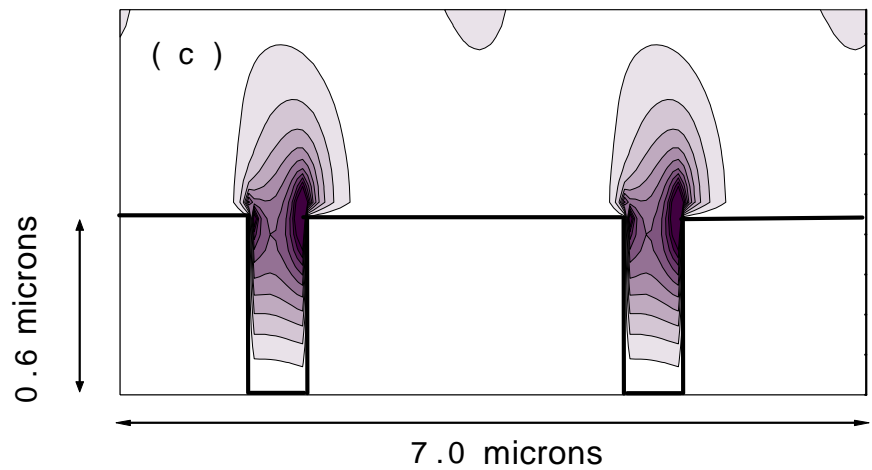
\end{document}